\documentclass[nofootinbib,prd]{revtex4}%
\usepackage{amsmath}
\usepackage{amsfonts}
\usepackage{amssymb}
\usepackage{graphicx}%
\usepackage{amsmath,amssymb}
\usepackage{mathrsfs}
\usepackage{graphicx}
\usepackage{color}
\usepackage{subfigure}
\usepackage{fancyhdr}
\usepackage{multirow}
\usepackage{float}
\usepackage{epsfig}
\usepackage{amsfonts}
\usepackage{bm}

\begin{document}
\title{Yukawa modification of Traversable Wormholes supported by Holographic Dark Energy}

\author{Remo Garattini}
\email{remo.garattini@unibg.it}
\affiliation{Universit\`a degli Studi di Bergamo, Dipartimento di Ingegneria e Scienze
Applicate, Viale Marconi 5, 24044 Dalmine (Bergamo) Italy and I.N.F.N.-
sezione di Milano, Milan, Italy.}

\author{Phongpichit Channuie}
\email{phongpichit.ch@mail.wu.ac.th}
\affiliation{College of Graduate Studies, Walailak University, Thasala, \\Nakhon Si Thammarat, 80160, Thailand\\School of Science, Walailak University, Thasala, \\Nakhon Si Thammarat, 80160, Thailand}

\author{Kirill Zatrimaylov}
\email{kirill.zatrimaylov@unibg.it}
\affiliation{Universit\`a degli Studi di Bergamo, Dipartimento di Ingegneria e Scienze
Applicate, Viale Marconi 5, 24044 Dalmine (Bergamo) Italy}

\begin{abstract}
Inspired by holographic dark energy models, we extend the analysis performed
in Ref.\cite{RGPC} taking into account Yukawa deformations on the different
energy density (HDE) profiles. The profiles we have explored are: the
Bekenstein-Hawking HDE, the Moradpour energy density, the Standard Renyi HDE
and the Mixed Energy Density. With the help of an inhomogeneous Equation of
State of the form $p_{r}(r)=\omega _{r}\left( r\right) \rho (r)$, we have
considered Zero Tidal Forces. Differently from the original case, only a
particular case of the family of the Bekenstein-Hawking HDE profiles
distorted by the Yukawa term has developed a divergent $\omega _{r}\left(
r\right) $ for $r\rightarrow \infty $. To cure such a divergence we have
introduced a modification at large distances in order to have a finite
result. This means that the Zero Tidal Forces can be imposed in a very large
but limited region of the spacetime. Such modification did not change the
behavior of the Equation of State close to the throat. For every proposal we
have computed the components of the Stress-Energy Tensor. What we have found
is that the Yukawa-Bekenstein-Hawking HDE is the only positive profile,
while the  Yukawa-Moradpour and the Yukawa-Renyi have a limited region where
the energy density is positive. The remaining Yukawa-Mixed Energy Density is
always negative. This means that the Yukawa distortion introduces a kind of
negative energy density. 

\end{abstract}
\maketitle

\section{Introduction}

Traversable wormholes (TWs) represent one of the most fascinating
predictions of General Relativity, providing hypothetical tunnels connecting
distant regions of spacetime. Since their modern formulation by \textit{%
Morris and Thorne} ~\cite{MT}, TWs have been extensively studied as
geometrical structures requiring the violation of classical energy
conditions, particularly the null energy condition (NEC), i.e., $\rho
+p_{r}<0$. Such violations imply the necessity of \textquotedblleft exotic
matter,\textquotedblright\ whose physical origin remains one of the central
challenges in wormhole physics. A first attempt to introduce such exotic
sources has been done by H.Ellis~\cite{Ellis} and K. Bronnikov~\cite{Bronnikov}%
, who considered scalar fields with wrong polarity. On this ground, attempts
to limit the usage of exotic matter has been performed with the introduction
of phantom fields~\cite{pWH1,pWH2,pWH3}, namely fields satisfying an Equation
of State (EoS) of the form $p_{r}=\omega \rho $ with $\omega <-1$ in such a
way the NEC violation becomes%
\begin{equation}
p_{r}+\rho =\rho \left( 1+\omega \right) <0
\end{equation}%
with $\rho >0$. Over the years, several candidates for exotic matter have
been proposed, ranging from quantum vacuum effects such as Casimir energy~\cite{MTY,Hochberg,Visser,EPJC} to phenomenological dark energy models. Among these, \textit{%
holographic dark energy} (HDE) has emerged as a particularly compelling
framework. Rooted in the holographic principle, HDE connects the ultraviolet
(UV) and infrared (IR) cutoffs of effective field theories, suggesting that
the vacuum energy density scales inversely with the square of a
characteristic length scale. In its standard formulation, the HDE density
takes the form~\cite{Li:2004rb}
\begin{equation}
\rho _{\text{HDE}}=3c^{2}M_{p}^{2}L^{-2},
\end{equation}%
where $L$ is typically associated with a cosmological horizon. This approach
has been widely used to address the dark energy problem and has also been
extended to compact objects and wormhole geometries~\cite{RGPC,Wang:2016pwd,Fatima:2024vag,Gonzalez-Diaz:2005toe,Paul:2025vem,Alshammari:2026wnf,Kiroriwal:2025trq,Chattopadhyay:2014jua,Paul:2025non,Debnath:2020aeg,Errehymy:2026nna,Nashed:2026rah,Chaudhary:2025khr,Rizwan:2026gxh,Nashed:2026vwv,Naseer:2026dxz,Chaudhary:2026uzs}.

Recent studies have shown that HDE-inspired energy densities can
successfully support traversable wormhole solutions, even when starting from
positive energy profiles \cite{RGPC,Nashed:2026vwv,Rizwan:2026gxh}. However, this often requires the introduction of
an inhomogeneous EoS, leading the system into the phantom regime near or
beyond the wormhole throat. Furthermore, while certain configurations allow
the imposition of zero tidal forces, they frequently result in pathological
behaviors such as divergences in the EoS parameter at large radial
distances. This limitation motivates the search for modified frameworks that
preserve desirable physical properties while avoiding such divergences. On a
different front, modifications of gravitational interactions at short or
intermediate scales have been widely investigated through Yukawa-type
corrections. Originally introduced by Yukawa to describe nuclear forces, the
Yukawa potential~\cite{Yukawa}%
\begin{equation}
V(r)=-\frac{\alpha }{r}e^{-\mu r},
\end{equation}%
represents a screened interaction with finite range $\mu ^{-1}$. When
applied to gravity, such corrections lead to a modified Newtonian potential
of the form 
\begin{equation}
V(r)=-\frac{Gm_{1}m_{2}}{r}\left( 1+\alpha e^{-\mu r}\right) ,
\end{equation}%
which naturally arises in various contexts including massive gravity~\cite%
{StarsGraviton}, modified gravity theories~\cite{MOG}, and phenomenological
models addressing galaxy rotation curves~\cite{MS}.

The incorporation of Yukawa corrections into wormhole physics has attracted
increasing attention~\cite{YC,YC1,YC2,YC3,YC4,YC5,YC6,YC7}. In particular, Yukawa-deformed
Casimir wormholes have demonstrated that the inclusion of exponential
suppression factors can regulate the behavior of the geometry, ensuring
asymptotic flatness and improving the physical viability of solutions. These
modifications influence the shape function, energy density, and pressure
profiles, often leading to improved control over the violation of energy
conditions and the asymptotic structure of spacetime~\cite{YC}.
Despite these advances, the interplay between Yukawa modifications and
holographic dark energy in supporting traversable wormholes remains largely
unexplored. This gap is particularly intriguing because both frameworks
address fundamental aspects of gravitational physics from different
perspectives: HDE emerges from quantum gravity considerations and the
holographic principle, while Yukawa corrections arise from effective field
theory and modified gravity scenarios. Combining these two approaches may
therefore provide a richer and more flexible framework for constructing
physically viable wormhole solutions.

In this work, we investigate traversable wormholes supported by holographic
dark energy under Yukawa-type modifications. Specifically, we introduce
Yukawa corrections into the source and examine their impact on the wormhole
structure. Our goal is to address key issues present in standard HDE
wormhole models, such as the divergence of the EoS parameter and the
limitations imposed by zero tidal force conditions. We analyze how Yukawa
screening affects the behavior of the shape function and the flaring-out
condition, the regularity and finiteness of the redshift function, the
asymptotic properties of spacetime, and the violation (or possible
mitigation) of the NEC. A particular emphasis is placed on constructing
models in which the Yukawa modification acts as a regulator, ensuring that
physically relevant quantities remain finite at large distances while
preserving the essential features of the wormhole near the throat. To
further proceed, we consider a static and spherically symmetric
Morris-Thorne traversable wormhole in Schwarzschild coordinates given by 
\cite{MT} 
\begin{equation}
ds^{2}=-e^{2\Phi (r)}dt^{2}+\frac{dr^{2}}{1-\frac{b(r)}{r}}+r^{2}\left(
d\theta ^{2}+\sin ^{2}\theta d\phi ^{2}\right) ,  \label{ds}
\end{equation}%
in which $\Phi (r)$ and $b(r)$ are the redshift and shape functions,
respectively. In the wormhole geometry, the redshift function $\Phi (r)$
should be finite in order to avoid the formation of an event horizon.
Moreover, the shape function $b(r)$ determines the wormhole geometry, with
the following condition $b(r_{0})=r_{0}$, in which $r_{0}$ is the radius of
the wormhole throat. Consequently, the shape function must satisfy the
flaring-out condition \cite{MT}: 
\begin{equation}
\frac{b(r)-rb^{\prime }(r)}{b^{2}(r)}>0,
\end{equation}%
in which $b^{\prime }(r_{0})<1$ must hold at the throat of the wormhole.
With the help of the line element $\left( \ref{ds}\right) $, we obtain the
following set of equations resulting from the energy-momentum components to
yield $\left( \kappa =8\pi G\right) $%
\begin{equation}
\frac{b^{\prime }(r)}{r^{2}}=\kappa \rho (r),  \label{rho}
\end{equation}%
\begin{equation}
\left[ 2\left( 1-\frac{b(r)}{r}\right) \frac{\Phi ^{\prime }(r)}{r}-\frac{%
b(r)}{r^{3}}\right] =\kappa p_{r}(r)  \label{pr}
\end{equation}%
and%
\begin{align}
& \Bigg\{\left( 1-\frac{b\left( r\right) }{r}\right) \left[ \Phi ^{\prime
\prime }(r)+\Phi ^{\prime }(r)\left( \Phi ^{\prime }(r)+\frac{1}{r}\right) %
\right]  \notag \\
& -\frac{b^{\prime }\left( r\right) r-b\left( r\right) }{2r^{2}}\left( \Phi
^{\prime }(r)+\frac{1}{r}\right) \Bigg\}=\kappa p_{t}(r).  \label{pt}
\end{align}%
We can complete the EFE with the expression of the conservation of the
Stress-Energy Tensor (SET) which can be written in the same orthonormal
reference frame%
\begin{equation}
p_{r}^{\prime }\left( r\right) =\frac{2}{r}\left( p_{t}\left( r\right)
-p_{r}\left( r\right) \right) -\left( \rho \left( r\right) +p_{r}\left(
r\right) \right) \Phi ^{\prime }(r).  \label{Tmn}
\end{equation}%
If Zero Tidal Forces (ZTF) are considered, we can impose the following inhomogeneous EoS%
\begin{equation}
p_{r}(r)=\omega _{r}\left( r\right) \rho (r)  \label{EoS1}
\end{equation}%
and write $\Phi (r)=0$. This is equivalent to impose%
\begin{equation}
\omega _{r}\left( r\right) =-\frac{b(r)}{rb^{\prime }(r)}.  \label{or(r)}
\end{equation}%
Solving the previous equation with respect to $b(r)$, one finds the well
know profile%
\begin{equation}
b(r)=r_{0}\,\exp \left[ -\int_{r_{0}}^{r}\,\frac{d\bar{r}}{\omega _{r}\left( 
\bar{r}\right) \bar{r}}\right] \,  \label{form}
\end{equation}%
which must be consistent with the result obtained by solving Eq.$\left( \ref%
{rho}\right) $. It is straightforward to see that some energy density
profiles produce a divergent $\omega _{r}\left( r\right) $ to keep the
validity of ZTF. For instance, Casimir wormholes\cite{EPJC} and
Yukawa-Casimir wormholes\cite{YC} fall in this case, while
the Ellis-Bronnikov TW does not. Indeed, the Ellis-Bronnikov is described by
the following shape function%
\begin{equation}
b(r)=\frac{r_{0}^{2}}{r}
\end{equation}%
and $\omega _{r}\left( r\right) =1$, while the Casimir wormhole has a shape
function%
\begin{equation}
b(r)=\frac{2}{3}r_{0}+\frac{r_{0}^{2}}{3r}.
\end{equation}%
and the associated $\omega _{r}\left( r\right) $ is such that%
\begin{equation}
\omega _{r}\left( r\right) =\left( \frac{2r}{r_{0}}+1\right) \underset{%
r\rightarrow \infty }{\rightarrow }\infty .
\end{equation}%
A similar behavior holds also for Yukawa-Casimir wormholes. In Ref.\cite%
{RGPC} several HDE profiles like the Bekenstein-Hawking HDE, the Moradpour
energy density, the Standard Renyi HDE and the Mixed Energy Density have
been explored and all of them have shown a divergent $\omega _{r}\left(
r\right) $ that has been taken under control modifying the strucuture of $%
\omega _{r}\left( r\right) $ itself. A similar procedure has been used also
in the context of TW powered by Dark Matter\cite{RGFLKZ}. In this paper we
shall apply such a procedure to the above mentioned HDE profiles distorted by
a Yukawa term. The paper is organized as follows. In Section \ref{ch2}, we
consider a Yukawa modification on the Bekenstein-Hawking HDE, in section \ref%
{ch3}, we compute some of the features of the Yukawa-Bekenstein-Hawking HDE TW, in section \ref%
{ch4}, we consider a Yukawa deformation on the Moradpour energy density, in
section \ref{ch5}, we consider a Yukawa deformation on the Standard Renyi
HDE, in section \ref{ch6}, we consider a Yukawa deformation on the Mixed
Energy Density. Finally, in Section \ref{ch7}, we summarize and conclude. Units in which $\hbar =c=k=1$ are used throughout the paper.

\section{Yukawa-Bekenstein-Hawking (YBH) HDE}

\label{ch2}We begin to consider the following energy density profile%
\begin{equation}
\rho _{1}(r)=\frac{C\pi }{r^{2}}\left( a+\left( 1-a\right) \exp (-\mu
(r-r_{0})\,\right) ,  \label{rhoYBH}
\end{equation}%
where $\mu $ is a positive mass scale to be identified later and $a\geq 0$.
When $a=1$ or $\mu =0$, we recover the original BH energy density profile
discussed in Ref.\cite{RGPC}. However, the case $a=1$ will not be discussed
because cuts off the damping exponential.  As we can see, in order to keep $%
\rho _{1}(r)>0$, we are forced to constrain the $a$ parameter into the
interval $0\leq a<1$. It is straightforward to compute
the shape function which is represented by%
\begin{equation}
b(r)=r_{0}+\beta a\left( r-r_{0}\right) +\frac{\beta }{\mu }\left(
a-1\right) \left( \exp (-\mu (r-r_{0}))-1\right) ,  \label{b(r)YBH}
\end{equation}%
where $\beta =\kappa C\pi >0$. It is interesting to observe that the flare
out condition%
\begin{equation}
b^{\prime }(r_{0})=\beta <1
\end{equation}%
is independent on $a$ and put a constraint on the constant $C$ in agreement
with what found in Ref.\cite{RGPC}. Note that when $a\neq 0$%
\begin{equation}
r\rightarrow \infty ,\,\qquad b(r)\rightarrow \beta ar,
\end{equation}%
while when%
\begin{equation}
\mu \rightarrow \infty ,\qquad b(r)\rightarrow r_{0}+\beta a\left(
r-r_{0}\right) .
\end{equation}%
On the other hand when $a=0$, one finds%
\begin{equation}
b(r)=r_{0}+\frac{\beta }{\mu }\left( 1-\exp (-\mu (r-r_{0})\right) 
\end{equation}%
and%
\begin{equation}
b(r)=r_{0}+\frac{\beta }{\mu }
\end{equation}%
for $r\rightarrow \infty $. If, in addition, we also consider the limit in
which $\mu \rightarrow \infty $, then%
\begin{equation}
b(r)=r_{0}.
\end{equation}%
To impose the Zero Tidal Forces (ZTF), we have to compute%
\begin{equation}
\omega _{r}\left( r\right) =-\frac{b(r)}{rb^{\prime }(r)}=\frac{\mu
r_{0}+\left( a-1\right) \beta \exp (-\mu (r-r_{0}))+\beta \left( 1+a(\mu
(r-r_{0})-1)\right) }{\mu \beta r\left( \left( a-1\right) \exp (-\mu
(r-r_{0}))-a\right) }.
\end{equation}%
At the throat one finds%
\begin{equation}
\omega _{r}\left( r_{0}\right) =-\frac{1}{\beta },  \label{or(r0)}
\end{equation}%
which is independent on $a$. On the other hand, when $r\rightarrow \infty $,
we discover that for $a=0$%
\begin{equation}
\omega _{r}\left( r\right) =\frac{\beta -\left( \mu r_{0}+\beta \right) \exp
(\mu (r-r_{0}))}{\mu \beta r}\rightarrow -\infty ,  \label{or(r)0}
\end{equation}%
while for $a\neq 0$, we obtain%
\begin{equation}
\omega _{r}\left( r\right) \underset{r\rightarrow \infty }{\rightarrow -1}+%
\frac{\beta a\mu r_{0}+\beta a-\mu r_{0}-\beta }{\mu a\beta r}+O\left( \exp
\left( -\mu r\right) \right) .
\end{equation}%
Therefore for $a\neq 0$,%
\begin{equation}
p_{r}\left( r\right) =-\frac{b(r)}{\kappa r^{3}}=-\frac{1}{\kappa r^{3}}%
\left( r_{0}+\beta a\left( r-r_{0}\right) +\frac{\beta }{\mu }\left(
a-1\right) \left( \exp (-\mu (r-r_{0}))-1\right) \right) 
\end{equation}%
and%
\begin{equation}
p_{t}\left( r\right) =\frac{b\!\left( r\right) -rb\!^{\prime }\left(
r\right) }{2\kappa r^{3}}=\frac{\beta \left( \mu r+1\right) \left(
a-1\right) e^{-\mu \left( r-r_{0}\right) }+\beta \left( 1-a\mu
r_{0}-a\right) +\mu r_{0}}{2\kappa r^{3}\mu }.
\end{equation}%
By focussing our attention only on the radial coefficient, we can observe
that%
\begin{equation}
\frac{dr^{2}}{\left( 1-\beta a\right) \left( 1-\frac{r_{0}}{r}\right) -\frac{%
\beta }{\mu r}\left( a-1\right) \left( \exp (-\mu (r-r_{0}))-1\right) }%
\qquad \underset{r\rightarrow \infty }{\longrightarrow }\qquad \frac{dr^{2}}{%
\left( 1-\beta a\right) \left( 1-\frac{r_{0}}{r}\right) +\frac{\beta }{\mu r}%
\left( a-1\right) }.  \label{ds21aa}
\end{equation}%
This means that, only for $a=1$ a \textit{Global Monopole} appears\cite%
{EPJC}. Even in this case, since $\beta <1$, we have an excess of the solid
angle for the line element $\left( \ref{ds}\right) $. Now, we study the case 
$a=0$. To take under control the divergence in Eq.$\left( \ref{or(r)0}%
\right) $, we can consider the following modification\cite{RGPC}%
\begin{equation}
\omega _{r}\left( r\right) =-\frac{b(r)f\left( r\right) }{rb^{\prime }(r)},
\label{oYBH}
\end{equation}%
where%
\begin{equation}
f\left( r\right) =\left\{ 
\begin{array}{cc}
1 & r_{0}\leq r\leq \bar{r} \\ 
\exp (-\mu (r-\bar{r})) & \bar{r}\leq r<\infty 
\end{array}%
\right. .
\end{equation}%
Thus%
\begin{equation}
\omega _{r}\left( r\right) =\left\{ 
\begin{array}{cc}
\frac{\beta \left( \exp (-\mu (r-r_{0})-1\right) -\mu r_{0}}{\beta \mu r\exp
(-\mu (r-r_{0}))} & r_{0}\leq r\leq \bar{r} \\ 
\frac{\beta \left( \exp (-\mu (r-r_{0})-1\right) -\mu r_{0}}{\beta \mu r\exp
(-\mu (\bar{r}-r_{0}))} & \bar{r}\leq r<\infty 
\end{array}%
\right. .
\end{equation}%
Thanks to the assumption $\left( \ref{oYBH}\right) $, the second EFE $\left( %
\ref{pr}\right) $ can be cast into the following form%
\begin{equation}
\Phi ^{\prime }(r)=\frac{b(r)}{2r}\frac{1-f\left( r\right) }{r-b(r)}.
\end{equation}%
The ZTF can be imposed only in the range $\left[ r_{0},\bar{r}\right] $,
while in the remaining part of the real axis, we can write%
\begin{equation}
\Phi ^{\prime }(r)=\frac{\left( \mu r_{0}+\beta -\beta e^{-\mu \left(
r-r_{0}\right) }\right) \left( 1-e^{-\mu \left( r-\bar{r}\right) }\right) }{%
2\left( \beta \left( e^{-\mu \left( r-r_{0}\right) }-1\right) +\mu \left(
r-r_{0}\right) \right) r}.
\end{equation}%
When $r\rightarrow \infty $, we can approximate the previous equation with%
\begin{equation}
\Phi ^{\prime }(r)\simeq \frac{\mu r_{0}+\beta }{2\mu r^{2}}+O\left( \frac{1%
}{r^{3}}\right) 
\end{equation}%
leading to%
\begin{equation}
\Phi (r)\simeq -\frac{\mu r_{0}+\beta }{2\mu r}.  \label{Phi'inf}
\end{equation}%
Before going on, we have to compute the radial pressure $p_{r}\left(
r\right) $ obtained with the help of Eq.$\left( \ref{oYBH}\right) $. We
obtain%
\begin{equation}
p_{r}\left( r\right) =-\frac{b(r)f\left( r\right) }{\kappa r^{3}}=\left\{ 
\begin{array}{cc}
-\frac{\mu r_{0}+\beta \left( 1-\exp (-\mu (r-r_{0}))\right) }{\kappa \mu
r^{3}} & r_{0}\leq r\leq \bar{r} \\ 
&  \\ 
-\frac{\mu r_{0}+\beta \left( 1-\exp (-\mu (r-r_{0}))\right) \exp (-\mu (r-%
\bar{r}))}{\kappa \mu r^{3}} & \bar{r}\leq r<\infty 
\end{array}%
\right. 
\end{equation}%
The last EFE\ to be computed is the third one leading to $p_{t}\left(
r\right) $. We obtain%
\begin{equation}
p_{t}\left( r\right) =\frac{\left( r\left( f\!\left( r\right) +1\right)
b\!\left( r\right) -2r^{2}f\!\left( r\right) \right) b\!^{\prime }\left(
r\right) -\left( 2r\left( r-b\!\left( r\right) \right) f^{\prime }\!\left(
r\right) -f\!\left( r\right) \left( 2r+\left( f\left( r\right) -3\right)
b\left( r\right) \right) \right) b\!\left( r\right) }{4\kappa r^{3}\left(
r-b\!\left( r\right) \right) }.
\end{equation}%
When $r_{0}\leq r\leq \bar{r}$, one gets%
\begin{equation}
p_{t}\left( r\right) =\frac{b\!\left( r\right) -rb\!^{\prime }\left(
r\right) }{2\kappa r^{3}}=\frac{\mu r_{0}+\beta -\beta \left( \mu r+1\right)
\exp (-\mu (r-r_{0}))}{2\kappa \mu r^{3}},
\end{equation}%
while when $\bar{r}\leq r<\infty $, we use the approximation $\left( \ref%
{Phi'inf}\right) $ to write%
\begin{equation}
p_{t}\left( r\right) \simeq \frac{\mu r_{0}+\beta }{2\kappa \mu r^{3}}%
+O\left( \frac{1}{r^{4}}\right) .
\end{equation}%
Contrary to the radial pressure, $p_{t}\left( r\right) $ exhibits a
discontinuity in $\bar{r}$. Indeed, one can write%
\begin{equation}
p_{t}\left( r\right) =\left\{ 
\begin{array}{cc}
\frac{\mu r_{0}+\beta -\beta \left( \mu \bar{r}+1\right) \exp (-\mu (\bar{r}%
-r_{0}))}{2\kappa \mu \bar{r}^{3}} & r\rightarrow \bar{r}_{-} \\ 
&  \\ 
\frac{\left( \mu \bar{r}+1\right) \left( \mu r_{0}+\beta \right) -\beta
\left( 2\mu \bar{r}+1\right) \exp (-\mu (r_{0}-\bar{r}))}{2\bar{r}^{3}\mu }
& r\rightarrow \bar{r}_{+}%
\end{array}%
.\right. 
\end{equation}%
Nevertheless the difference between the left and the right limit is%
\begin{equation}
p_{t}^{-}\left( r\right) -p_{t}^{+}\left( r\right) =\frac{\beta \exp (-\mu
(r_{0}-\bar{r}))}{2\bar{r}^{2}}-\frac{\mu r_{0}}{2\bar{r}^{2}}-\frac{\beta }{%
2\bar{r}^{2}}
\end{equation}%
showing that the difference becomes smaller and smaller whenever $\bar{r}$
becomes larger and larger.


\section{Features of the YBH-HDE TW}

\label{ch3}In section \ref{ch2}, we have computed the shape function $\left( %
\ref{b(r)YBH}\right) $ obtained by the YBH-HDE energy density. Here we want
to discuss some of its properties, the first of which is the proper radial
distance which is related to the shape function by the following definition%
\begin{equation}
l\left( r\right) =\pm \int_{r_{0}}^{r}\frac{dr^{\prime }}{\sqrt{1-\frac{%
b\left( r^{\prime }\right) }{r^{\prime }}}}.
\end{equation}%
Due to the complexity of the shape function we proceed to evaluate the
proper length in two regions: close to the throat and very far from the
throat. We obtain%
\begin{equation}
l\left( r\right) \simeq \left\{ 
\begin{array}{ll}
\pm \left( \frac{r}{\sqrt{1-\beta a}}-\frac{\left( a\beta \mu r_{0}+a\beta
-\mu r_{0}-\beta \right) }{2\mu \sqrt{\left( 1-\beta a\right) ^{3}}}\ln
\left( r\right) +O\left( \frac{1}{r}\right) \right)  & r\rightarrow \infty 
\\ 
\pm 2\sqrt{r_{0}\left( r-r_{0}\right) }/\sqrt{1-\beta }+O\left( \left(
r-r_{0}\right) ^{\frac{3}{2}}\right)  & r\rightarrow r_{0}%
\end{array}%
\right. ,
\end{equation}%

where the\textquotedblleft $\pm $\textquotedblright\ depends on the wormhole
side we are. The proper radial distance is an essential tool to estimate the
possible time trip in going from one station located in the lower universe,
say at $l=-l_{1}$, and ending up in the upper universe station, say at $%
l=l_{2}$. Following Ref.\cite{MT}, we shall locate $l_{1}$ and $l_{2}$ at a
value of the radius such that $l_{1}\simeq l_{2}\simeq 10^{4}r_{0}$ that it
means $1-b\left( r\right) /r\simeq 1$. Assume that the traveller has a
radial velocity $v\left( r\right) $, as measured by a static observer
positioned at $r$. One may relate the proper distance travelled $dl$, radius
travelled $dr$, coordinate time lapse $dt$, and proper time lapse as
measured by the observer $d\tau $, by the following relationships%
\begin{equation}
v=e^{-\Phi \left( r\right) }\frac{dl}{dt}=e^{-\Phi \left( r\right) }\left( 1-%
\frac{b\left( r\right) }{r}\right) ^{-\frac{1}{2}}\frac{dr}{dt}
\end{equation}%
and%
\begin{equation}
v\gamma =\frac{dl}{d\tau }=\mp \left( 1-\frac{b\left( r\right) }{r}\right)
^{-\frac{1}{2}}\frac{dr}{d\tau };\qquad \gamma =\left( 1-\frac{v^{2}\left(
r\right) }{c^{2}}\right) ^{-\frac{1}{2}}
\end{equation}%
respectively. If the traveler journeys with constant speed $v$ and $\gamma
\simeq 1$, then the total time is given by%
\begin{equation}
\Delta t=\int_{r_{0}}^{r}\frac{e^{-\Phi \left( r^{\prime }\right)
}dr^{\prime }}{v\sqrt{1-\frac{b\left( r^{\prime }\right) }{r^{\prime }}}},
\end{equation}%
while the proper total time is defined as%
\begin{equation}
\Delta \tau =\int_{r_{0}}^{r}\frac{dr^{\prime }}{v\sqrt{1-\frac{b\left(
r^{\prime }\right) }{r^{\prime }}}}.
\end{equation}%
As we can see%
\begin{equation}
\Delta \tau =\Delta t=\frac{r}{v\sqrt{1-\beta a}}
\end{equation}%
but only in the range $\left[ r_{0},\bar{r}\right] $. Outside of this range,
it appears that $\Delta \tau \neq \Delta t$. However from Eq.$\left( \ref%
{Phi'inf}\right) $, one can say that $\Phi \left( r\right) \rightarrow 0$,
for $r\rightarrow \infty $. This means that even in this case we can claim
that $\Delta \tau \simeq \Delta t$. On the same ground, we can compute the
embedded surface, which is defined by%
\begin{equation}
z\left( r\right) =\pm \int_{r_{0}}^{r}\frac{dr^{\prime }}{\sqrt{\frac{%
r^{\prime }}{b\left( r^{\prime }\right) }-1}}
\end{equation}%
and, in the present case, we find%
\begin{equation}
z\left( r\right) \simeq \pm \frac{2\sqrt{r_{0}}\sqrt{r-r_{0}}}{\sqrt{1-\beta 
}}+O\left( \left( r-r_{0}\right) ^{\frac{3}{2}}\right) 
\end{equation}%
To further investigate the properties of the shape function $\left( \ref%
{b(r)YBH}\right) $, we consider the computation of the total gravitational
energy for a wormhole\cite{NZCP}, defined as%
\begin{gather}
E_{G}\left( r\right) =\int_{r_{0}}^{r}\left[ 1-\sqrt{\frac{1}{1-b\left(
r^{\prime }\right) /r^{\prime }}}\right] \rho \left( r^{\prime }\right)
dr^{\prime }r^{\prime 2}  \notag \\
+\frac{r_{0}}{2G}=M-M_{\pm }^{P},
\end{gather}%
where $M$ is the total mass $M$ and $M^{P}$ is the proper mass,
respectively. Even in this case, the \textquotedblleft $\pm $%
\textquotedblright\ depends one the wormhole side we are. For the total
mass, one finds%
\begin{equation}
M=\int_{r_{0}}^{r}4\pi \rho \left( r^{\prime }\right) r^{\prime 2}dr^{\prime
}+\frac{r_{0}}{2G}=\frac{4\pi }{\kappa }\left( b\left( r\right)
-r_{0}\right) +\frac{r_{0}}{2G}=\frac{b\left( r\right) }{2G},
\end{equation}%
while for the proper mass, one gets%
\begin{equation}
M_{\pm }^{P}=\pm \int_{r_{0}}^{r}\frac{4\pi \rho \left( r^{\prime }\right)
r^{\prime 2}}{\sqrt{1-b\left( r^{\prime }\right) /r^{\prime }}}dr^{\prime }.
\end{equation}%
Therefore, the general result is%
\begin{eqnarray}
E_{G}\left( r\right)  &=&\frac{b\left( r\right) }{2G}\pm \int_{r_{0}}^{r}%
\frac{4\pi \rho \left( r^{\prime }\right) r^{\prime 2}}{\sqrt{1-b\left(
r^{\prime }\right) /r^{\prime }}}dr^{\prime }  \notag \\
&=&\frac{b\left( r\right) }{2G}\pm \frac{4\pi }{\kappa }\int_{r_{0}}^{r}%
\frac{b^{\prime }\left( r^{\prime }\right) }{\sqrt{1-b\left( r^{\prime
}\right) /r^{\prime }}}dr^{\prime },
\end{eqnarray}%
where we have used the first EFE to eliminate $\rho \left( r^{\prime
}\right) $. We have to distinguish the case in which $a\neq 0$ to the case
where $a=0$. When $a\neq 0$ and $r\rightarrow \infty $ the total mass $M$ is%
\begin{equation}
M\underset{r\rightarrow \infty }{\simeq }\frac{r_{0}+\beta a\left(
r-r_{0}\right) -\frac{\beta }{\mu }\left( a-1\right) }{2G}\underset{%
r\rightarrow \infty }{\simeq }\frac{\beta ar}{2G},
\end{equation}%
while the proper mass $M_{\pm }^{P}$ can be written as%
\begin{equation}
M_{\pm }^{P}=\pm \frac{4\pi }{\kappa }\int_{r_{0}}^{r}\frac{b^{\prime
}\left( r^{\prime }\right) }{\sqrt{1-b\left( r^{\prime }\right) /r^{\prime }}%
}dr^{\prime }\underset{r\rightarrow \infty }{\simeq }\pm \frac{\beta ar}{2G%
\sqrt{1-\beta a}}.
\end{equation}%
On the other hand when $r\rightarrow r_{0}$, we can write for $M_{\pm }^{P}$%
\begin{equation}
M_{\pm }^{P}\underset{r\rightarrow r_{0}}{\simeq }\pm \frac{\beta \sqrt{%
r_{0}\left( r-r_{0}\right) }}{2G\sqrt{1-\beta }}.
\end{equation}%
In summary for $a\neq 0$, we can write%
\begin{equation}
E_{G}\left( r\right) =\left\{ 
\begin{array}{c}
M-M_{\pm }^{P}\underset{r\rightarrow \infty }{\simeq }\frac{\beta ar}{2G}%
\left( 1\pm \frac{1}{\sqrt{1-\beta a}}\right)  \\ 
\\ 
M-M_{\pm }^{P}\underset{r\rightarrow r_{0}}{\simeq }\frac{1}{2G}\left(
r_{0}\pm \frac{\beta \sqrt{r_{0}\left( r-r_{0}\right) }}{\sqrt{1-\beta }}%
\right) .%
\end{array}%
\right. 
\end{equation}%
Now we are going to consider the case in which $a=0$ and $r\rightarrow
\infty $. For the total mass, one obtains%
\begin{equation}
M\underset{r\rightarrow \infty }{\simeq }\frac{1}{2G}\left( r_{0}+\frac{%
\beta }{\mu }\right) ,
\end{equation}%
while the proper mass $M_{\pm }^{P}$ is vanishing because of the Yukawa term
on the energy density. For $r\rightarrow r_{0}$, the total mass is simply%
\begin{equation}
M\underset{r\rightarrow r_{0}}{\simeq }\frac{r_{0}}{2G},
\end{equation}%
while the proper mass becomes%
\begin{equation}
M_{\pm }^{P}=\pm \frac{4\pi }{\kappa }\int_{r_{0}}^{r}\frac{b^{\prime
}\left( r^{\prime }\right) }{\sqrt{1-b\left( r^{\prime }\right) /r^{\prime }}%
}dr^{\prime }\underset{r\rightarrow r_{0}}{\simeq }\pm \frac{\beta \sqrt{%
r_{0}\left( r-r_{0}\right) }}{2G\sqrt{1-\beta }}.
\end{equation}%
In summary for $a=0$, we can write%
\begin{equation}
E_{G}\left( r\right) =\left\{ 
\begin{array}{c}
M-M_{\pm }^{P}\underset{r\rightarrow \infty }{\simeq }\frac{1}{2G}\left(
r_{0}+\frac{\beta }{\mu }\right)  \\ 
\\ 
M-M_{\pm }^{P}\underset{r\rightarrow r_{0}}{\simeq }\frac{1}{2G}\left(
r_{0}\pm \frac{\beta \sqrt{r_{0}\left( r-r_{0}\right) }}{\sqrt{1-\beta }}%
\right) .%
\end{array}%
\right. 
\end{equation}%
It appears that, close to the throat, $E_{G}\left( r\right) $ behaves in the
same way, while at infinity, only for $a=0$, one obtains a finite value.
Even if we restrict the range to the interval $\left[ r_{0},\bar{r}\right] $%
, $E_{G}\left( \bar{r},a=0\right) \neq E_{G}\left( \bar{r},a\neq 0\right) $.
An important traversability condition is that the acceleration felt by the
traveller should not exceed Earth's gravity $g_{\oplus }\simeq 980$ $cm/s^{2}
$. In an orthonormal basis of the traveller's proper reference frame, we can
find%
\begin{equation}
\left\vert \mathbf{a}\right\vert =\left\vert \sqrt{1-\frac{b\left( r\right) 
}{r}}e^{-\Phi \left( r\right) }\left( \gamma e^{\Phi \left( r\right)
}\right) ^{\prime }\right\vert \leq \frac{g_{\oplus }}{c^{2}}.
\end{equation}%
If we assume a constant speed and $\gamma \simeq 1$, then we can deduce that 
$\left\vert \mathbf{a}\right\vert \rightarrow 0$, especially in proximity of
the throat. Always following Ref.\cite{MT}, we can estimate the tidal forces
by imposing an upper bound represented by $g_{\oplus }$. The radial tidal
constraint 
\begin{gather}
\left\vert \left( 1-\frac{b\left( r\right) }{r}\right) \left[ \Phi ^{\prime
\prime }\left( r\right) +\left( \Phi ^{\prime }\left( r\right) \right)
^{2}\right. \right.   \notag \\
\left. \left. -\frac{b^{\prime }\left( r\right) r-b\left( r\right) }{%
2r\left( r-b\left( r\right) \right) }\Phi ^{\prime }\left( r\right) \right]
c^{2}\right\vert \left\vert \eta ^{\hat{1}^{\prime }}\right\vert \leq
g_{\oplus },  \label{RTC}
\end{gather}%
constrains the redshift function, and the lateral tidal constraint%
\begin{gather}
\left\vert \frac{\gamma ^{2}c^{2}}{2r^{2}}\left[ \frac{v^{2}\left( r\right) 
}{c^{2}}\left( b^{\prime }\left( r\right) -\frac{b\left( r\right) }{r}%
\right) \right. \right.   \notag \\
\left. \left. +2r\left( r-b\left( r\right) \right) \Phi ^{\prime }\left(
r\right) \right] \right\vert \left\vert \eta ^{\hat{2}^{\prime }}\right\vert
\leq g_{\oplus },  \label{LTC}
\end{gather}%
constrains the velocity with which observers traverse the wormhole. $\eta ^{%
\hat{1}^{\prime }}$ and $\eta ^{\hat{2}^{\prime }}$ represent the size of
the traveller. In Ref.\cite{MT}, they are fixed approximately equal, at the
symbolic value of $2$ $m$. We can restrict our evaluation on the range $%
\left[ r_{0},\bar{r}\right] $ without loss of generality provided we assume
that the external boundary $\bar{r}=10^{4}r_{0}\simeq $ $l_{1}\simeq l_{2}$,
namely the location of the stations. Therefore the inequality $\left( \ref%
{RTC}\right) $ is always satisfied, while for the lateral tidal contraint,
we find%
\begin{equation}
\frac{v^{2}}{r_{0}^{2}}\left\vert \beta -1\right\vert \lesssim g_{\oplus
}\qquad \Longrightarrow \qquad v\lesssim r_{0}\sqrt{\frac{g_{\oplus }}{%
\left\vert \beta -1\right\vert }}.  \label{LTCt}
\end{equation}%
If the observer has a vanishing $v$, then the tidal forces are null. We can
use these last estimates to complete the evaluation of the crossing time
which approximately is%
\begin{equation}
\Delta t\simeq \frac{2\times 10^{4}r_{0}}{v\sqrt{1-\beta a}}\simeq \frac{%
2\times 10^{4}\sqrt{\left\vert \beta -1\right\vert }}{\sqrt{g_{\oplus }}%
\sqrt{1-\beta a}}\simeq \frac{2\times 10^{3}\sqrt{\left\vert \beta
-1\right\vert }}{\sqrt{1-\beta a}}s,
\end{equation}%
which is in agreement with the estimates found in Ref.\cite{MT}. Note that
there exists a dependence on the parameter $a$ and on the parameter $\beta $
connected to the initial BH-HDE energy density. Note also that for the
special value of $a=1$, such a dependence disappears. The last property we
are going to discuss is the \textquotedblleft total
amount\textquotedblright\ of ANEC violating matter in the spacetime~\cite{VKD}
which is described by%
\begin{equation}
I_{V}=\int [\rho (r)+p_{r}(r)]dV.
\end{equation}%
Changing the measure $dV$ into $r^{2}dr$, we find%
\begin{equation}
I_{V}=\frac{1}{\kappa }\int \left( r-b\left( r\right) \right) \left[ \ln
\left( \frac{e^{2\Phi (r)}}{1-\frac{b\left( r\right) }{r}}\right) \right]
^{\prime }dr  \label{IV}
\end{equation}%
and with the help of the shape funtion $\left( \ref{b(r)YBH}\right) $, one
gets%
\begin{equation}
I_{V}=\frac{1}{\kappa }\int_{r_{0}}^{\bar{r}}\frac{\left( a\mu
r_{0}+a-1\right) \beta -\mu r_{0}-\beta \left( \mu r+1\right) \left(
a-1\right) e^{-\mu \left( r-r_{0}\right) }}{\mu r}dr,
\end{equation}%
where we have used the fact that $\Phi \left( r\right) =0$ in the range $%
\left[ r_{0},\bar{r}\right] $. Close to the throat, the integrand function
becomes%
\begin{equation}
\left( \beta -1\right) +O\left( r-r_{0}\right) ,
\end{equation}%
this means that the ANEC can be arbitrarily small. For very large $r$, the
integrand function becomes%
\begin{equation}
\frac{\left( a\mu r_{0}+a-1\right) \beta -\mu r_{0}}{\mu r}
\end{equation}%
and after having integrated, we can write%
\begin{equation}
I_{V}\simeq \frac{\left( a\mu r_{0}+a-1\right) \beta -\mu r_{0}}{\kappa }\ln
\left( \frac{\bar{r}}{r}\right) .  \label{IVL}
\end{equation}%
Therefore we can conclude that even close to the external boundary the ANEC
can be considered arbitrarily small.

\section{Yukawa deformation on the Moradpour energy density}
 
\label{ch4}In this section, we are going to examine a Yukawa deformation of
the Moradpour profile~\cite{Manoharan:2022qll}
\begin{equation}
\rho\left(  r\right)  =\frac{C}{4\pi r^{2}\left(  \pi\lambda r^{2}+1\right)
},\label{rhoM}%
\end{equation}
with the introduction of the exponential term%
\begin{equation}
\rho_{YM}\left(  r\right)  =\frac{C\exp\left(  -\mu\left(  r-r_{0}\right)
\right)  }{4\pi r^{2}\left(  \pi\lambda r^{2}+1\right)  }.\label{MY}
\end{equation}
Plugging $\rho_{YM}\left(  r\right)  $ into the first EFE $\left(
\ref{rho}\right)  $ leads to the following shape function%
\begin{equation}
b\left(  r\right)  =r_{0}+\frac{C\kappa}{4\pi}\int_{r_{0}}^{r}\frac
{\exp\left(  -\mu\left(  r^{\prime}-r_{0}\right)  \right)  }{\pi\lambda
r^{\prime2}+1}dr^{\prime}.
\end{equation}
Despite of the simple form of the integrand, its evaluation is far to be immediate. Indeed, if we integrate by parts the term
\begin{gather}
\int_{r_{0}}^{r}\frac{\exp\left(  -\mu\left(  r^{\prime}-r_{0}\right)
\right)  }{\pi\lambda r^{\prime2}+1}dr^{\prime},
\end{gather}
we obtain
\begin{gather}
\left[  \frac{\tan^{-1}\left(  \sqrt{\pi\lambda}r^{\prime}\right)  }%
{\sqrt{\pi\lambda}}\exp\left(  -\mu\left(  r^{\prime}-r_{0}\right)  \right)
\right]  _{r_{0}}^{r}+\mu\int_{r_{0}}^{r}\frac{\tan^{-1}\left(  \sqrt
{\pi\lambda}r\right)  }{\sqrt{\pi\lambda}}\exp\left(  -\mu\left(  r^{\prime
}-r_{0}\right)  \right)  dr^{\prime},
\end{gather}
leading to the conclusion that the original integral cannot be solved in a closed form. Therefore, differently from the previous section, including a damping function of Yukawa type is not so fruitful. However, one can slightly change the strategy adopting the following profile $\left(  \ref{MY}\right)  $,
\begin{align}
\rho_{2}\left(  r\right)   &  =\frac{C}{4\pi r^{2}}\frac{d}{dr}\left(
\frac{\tan^{-1}\left(  \sqrt{\pi\lambda}r\right)  }{\sqrt{\pi\lambda}}%
\exp\left(  -r_{0}^{2}\lambda\mu\left(  r-r_{0}\right)  \right)  \right)
\nonumber\\
&  =\frac{C\exp\left(  -\mu\lambda r_{0}^{2}\left(  r-r_{0}\right)  \right)
}{4\pi r^{2}}\left(  \frac{1}{\pi\lambda r^{2}+1}-\mu\lambda r_{0}^{2}%
\frac{\tan^{-1}\left(  \sqrt{\pi\lambda}r\right)  }{\sqrt{\pi\lambda}}\right)
.\label{rhoM1}%
\end{align}
$\rho_{2}\left(  r\right)$ reduces to the integrand function associated to the profile $\left(
\ref{rhoM}\right)  $ when $\mu\rightarrow0$. We observe that when
$\mu\rightarrow\infty$, $\rho_{2}\left(  r\right)  \rightarrow0$ as well as
for $\lambda\rightarrow\infty$. For this asymptotic cases $b(r)=r_{0}$.
Moreover when $\lambda\rightarrow0$, we obtain the energy density described by
Eq.$\left(  \ref{rhoYBH}\right)  $. With the assumption $\left(
\ref{rhoM1}\right)  $, the shape function can be represented in a closed form,
namely%
\begin{equation}
b\left(  r\right)  =r_{0}+\frac{C\kappa}{4\pi\sqrt{\pi\lambda}}\left[
\tan^{-1}\left(  \sqrt{\pi\lambda}r\right)  \exp\left(  -\mu\lambda r_{0}%
^{2}\left(  r-r_{0}\right)  \right)  -\tan^{-1}\left(  \sqrt{\pi\lambda}%
r_{0}\right)  \right]  .\label{b(r)YM}%
\end{equation}
Note that $C$ and $\lambda$ have dimensions $\left[  L^{-2}\right]  $, while
$\mu$ has dimensions $\left[  L^{-1}\right]  $. The shape function $\left(
\ref{b(r)YM}\right)  $ is such that%
\begin{equation}
b(r)=\underset{r\rightarrow\infty}{\rightarrow}r_{0}-\frac{\kappa C}{4\pi
\sqrt{\pi\lambda}}\tan^{-1}\left(  \sqrt{\pi\lambda}\,r_{0}\right)
=b_{M_{Y},\infty},\label{b(r)2a}%
\end{equation}
which reduces to the value $r_{0}$ when $\lambda\rightarrow\infty$, as it
should be. To avoid a negative $b_{M_{Y},\infty}$, one can observe that the
unknown quantity $C$ must be bounded by%
\begin{equation}
\frac{4\pi\sqrt{\pi\lambda}r_{0}}{\kappa\tan^{-1}\left(  \sqrt{\pi\lambda
}\,r_{0}\right)  }\geq C.\label{bound}%
\end{equation}
The case in which the previous bound is saturated, then one gets
$b_{M_{Y},\infty}=0$. For the wormhole be traversable, we have to impose the
flare-out condition, represented by
\begin{equation}
b^{\prime}(r_{0})=\frac{C\kappa}{4\pi\sqrt{\pi\lambda}}\left[  \frac{\sqrt
{\pi\lambda}}{\pi\lambda r_{0}^{2}+1}-\mu\tan^{-1}\left(  \sqrt{\pi\lambda
}r_{0}\right)  \right]  <1.\label{b'(r0)M}%
\end{equation}
Substituting the saturated value of the bound $\left(  \ref{bound}\right)  $
into $\left(  \ref{b'(r0)M}\right)  $, we obtain%
\begin{equation}
b^{\prime}(r_{0})=\left[  \frac{r_{0}\sqrt{\pi\lambda}}{\tan^{-1}\left(
\sqrt{\pi\lambda}\,r_{0}\right)  \left(  \pi\lambda r_{0}^{2}+1\right)  }%
-\mu\lambda r_{0}^{3}\right]  <1
\end{equation}
which is always satisfied except for the limit in which $\lambda\rightarrow0$,
where $b^{\prime}(r_{0})=1$. To gain enough information on the $\Phi\left(
r\right)  $, we examine the original inhomogeneous EoS to see if a
modification is necessary. Such an EoS, if satisfied, allows us to impose ZTF
and set $\Phi\!\left(  r\right)  =0$ everywhere. To this purpose, we need
to compute Eq.$\left( \ref{or(r)}\right) $ which is represented by 
\begin{equation}
\omega _{r}\left( r\right) =\frac{\omega _{r}^{N}\left( r\right) }{\kappa
Cr\left( \sqrt{\pi \lambda }-\mu \lambda r_{0}^{2}\left( \pi \lambda
r^{2}+1\right) \tan ^{-1}\!\left( \sqrt{\pi \lambda }r\right) \,\right) },
\label{or(r)MY}
\end{equation}%
where%
\begin{equation}
\omega _{r}^{N}\left( r\right) =\left( \pi \lambda r^{2}+1\right) \left(
\left( \tan ^{-1}\!\left( \sqrt{\pi \lambda }r_{0}\right) C\kappa -4r_{0}\pi
^{\frac{3}{2}}\sqrt{\lambda }\right) \exp \left( \mu \lambda r_{0}^{2}\left(
r-r_{0}\right) \right) -\kappa C\tan ^{-1}\!\left( \sqrt{\pi \lambda }%
r\right) \right) .
\end{equation}%
We can observe that for $r\rightarrow \infty $, $\omega _{r}^{N}\left(
r\right) \rightarrow \infty $ and therefore also $\omega _{r}\left( r\right) 
$. To avoid the appearance of a divergence, we impose that%
\begin{equation}
C=\frac{4r_{0}\pi^{\frac{3}{2}}\sqrt{\lambda}}{\kappa\tan^{-1}\!\left(
\sqrt{\pi\lambda}r_{0}\right)  },\label{C}%
\end{equation}
which is coincident with the saturated bound $\left(  \ref{bound}\right)  $
and $\omega_{r}\left(  r\right)  $ becomes%
\begin{equation}
\omega_{r}\left(  r\right)  =\frac{\tan^{-1}\!\left(  \sqrt{\pi\lambda
}r\right)  \left(  \pi\lambda r^{2}+1\right)  }{r\left(  \mu\lambda r_{0}%
^{2}\left(  \pi\lambda r^{2}+1\right)  \tan^{-1}\!\left(  \sqrt{\pi\lambda
}r\right)  -\sqrt{\pi\lambda}\,\right)  }.\label{or(r)MYF}%
\end{equation}
Now, when $r\rightarrow\infty$%
\begin{equation}
\omega_{r}\left(  r\right)  =\frac{1}{\mu\lambda r_{0}^{2}r}+O\left(  \frac
{1}{r^{3}}\right)
\end{equation}
and, on the throat, we find%
\begin{equation}
\omega_{r}\left(  r_{0}\right)  =\frac{\tan^{-1}\!\left(  \sqrt{\pi\lambda
}r_{0}\right)  \left(  \pi\lambda r_{0}^{2}+1\right)  }{r_{0}\left(
\mu\lambda r_{0}^{2}\left(  \pi\lambda r_{0}^{2}+1\right)  \tan^{-1}\!\left(
\sqrt{\pi\lambda}r_{0}\right)  -\sqrt{\pi\lambda}\,\right)  }.\label{or(r)Mr0}%
\end{equation}
Note that with the constraint $\left(  \ref{C}\right)  $, the shape function
becomes%
\begin{align}
b\left(  r\right)   &  =r_{0}+\frac{r_{0}}{\tan^{-1}\!\left(  \sqrt{\pi
\lambda}r_{0}\right)  }\left[  \tan^{-1}\left(  \sqrt{\pi\lambda}r\right)
\exp\left(  -\mu\lambda r_{0}^{2}\left(  r-r_{0}\right)  \right)  -\tan
^{-1}\left(  \sqrt{\pi\lambda}r_{0}\right)  \right]  \nonumber\\
&  =r_{0}\frac{\tan^{-1}\left(  \sqrt{\pi\lambda}r\right)  \exp\left(
-\mu\lambda r_{0}^{2}\left(  r-r_{0}\right)  \right)  }{\tan^{-1}\!\left(
\sqrt{\pi\lambda}r_{0}\right)  }.\label{b(r)MC}%
\end{align}
To further proceed, it is convenient to adopt the profile $\left(
\ref{b(r)MC}\right)  $ to discuss the following extreme cases:

\begin{description}
\item[a)] $\lambda\rightarrow0$ and $0<\mu<+\infty$,

\item[b)] $\lambda\rightarrow+\infty$ and $0<\mu<+\infty$,

\item[c)] $\lambda\rightarrow+\infty$ and $\mu\rightarrow+\infty$,

\item[d)] $\lambda\rightarrow0$ and $\mu\rightarrow+\infty$ with $\lambda\mu$ finite.
\end{description}

For case \textbf{a)}, we find $b\left(  r\right)  =r$ which will be discarded
because it is not a physical solution. The same happens for the case
\textbf{b)}, where $b\left(  r\right)  \rightarrow0$. A similar behavior
appears also for the case \textbf{c}) . For case \textbf{d)}, we find that the
product $\mu\lambda$ is finite and
\begin{equation}
b\left(  r\right)  =r\exp\left(  -\mu\lambda r_{0}^{2}\left(  r-r_{0}\right)
\right)  .
\end{equation}
Case \textbf{d)} represents a TW with a Yukawa profile. The only unpleasant
feature is that the energy density is completely negative and not positive.
Therefore, even this last case will be discarded. Notwithstanding the profile
$\left(  \ref{b(r)MC}\right)  $ can be considered as a valid representation
for a TW powered by a Moradpour energy density source deformed by a Yukawa
term. The remaining components of the SET are:

the radial pressure%
\begin{equation}
p_{r}\left(  r\right)  =-\frac{b(r)}{\kappa r^{3}}=-\frac{r_{0}\tan
^{-1}\left(  \sqrt{\pi\lambda}r\right)  \exp\left(  -\mu\lambda r_{0}%
^{2}\left(  r-r_{0}\right)  \right)  }{\kappa r^{3}\tan^{-1}\!\left(
\sqrt{\pi\lambda}r_{0}\right)  }%
\end{equation}
and the tangential pressure%
\begin{equation}
p_{t}\left(  r\right)  =\frac{b\left(  r\right)  -b^{\prime}\left(  r\right)
r}{2\kappa r^{3}}=\frac{r_{0}\left(  \left(  r_{0}^{2}\lambda\mu r+1\right)
\left(  \pi\lambda r^{2}+1\right)  \tan^{-1}\left(  \sqrt{\pi\lambda}r\right)
-\sqrt{\pi\lambda}\,r\right)  }{2r^{3}\left(  \pi\lambda r^{2}+1\right)
\tan^{-1}\left(  \sqrt{\pi\lambda}r_{0}\right)  }\exp\left(  -\mu\lambda
r_{0}^{2}\left(  r-r_{0}\right)  \right)  .
\end{equation}

\textbf{Remark} Apparently our Yukawa corrected Moradpour energy density
$\left(  \ref{rhoM1}\right)  $ seems to avoid the introduction of negative
energy density. Unfortunately this is not the case as shown in the following
table%
\begin{equation}%
\begin{array}
[c]{cccc}%
x=r/r_{0} & y=\lambda r_{0}^{2} & z=\mu r_{0} & \rho_{2}\left(  r\right)  >0\\
- & 2 & 2 & -\\
- & 1 & 1 & -\\
1<x<1.517 & 1/2 & 1/2 & \surd\\
1<x<3.602 & 1/4 & 1/4 & \surd\\
1<x<26.04 & 1/20 & 1/20 & \surd\\
1<x<2.15 & 1/2 & 1/4 & \surd\\
1<x<2.56 & 1/4 & 1/2 & \surd
\end{array}
.
\end{equation}

\section{Yukawa deformation on the Standard Renyi HDE}

\label{ch5}In this section, we will consider the Renyi HDE
density~\cite{Manoharan:2022qll}
\begin{equation}
\rho\left(  r\right)  =\frac{C}{\lambda r^{4}}\ln\left(  1+\pi\lambda
r^{2}\right)  , \label{rho3}%
\end{equation}
modified by a Yukawa term like in section \ref{ch5}. In order to have a
solution in a closed form, we propose the following modification%
\begin{equation}
\rho_{3}\left(  r\right)  =\frac{1}{r^{2}}\frac{d}{dr}\left(  b_{3}\left(
r\right)  \exp\left(  -\mu\left(  r-r_{0}\right)  \right)  \right)  ,
\end{equation}
where $b_{3}\left(  r\right)  $ is such that%
\begin{equation}
\frac{d}{dr}b_{3}\left(  r\right)  =\frac{C}{\lambda r^{2}}\ln\left(
1+\pi\lambda r^{2}\right)  .
\end{equation}
Note that $C$ and $\lambda$ have dimensions $\left[  L^{-2}\right]  $ and are
positive. $\rho_{3}\left(  r\right)  $ vanishes for $r\rightarrow\infty$, as
well as for $\lambda\rightarrow\infty$ and $\mu\rightarrow\infty$. On the
other hand for $\lambda\rightarrow0$, $\rho_{3}\left(  r\right)  \rightarrow$
$\rho_{1}\left(  r\right)  $ which is described by Eq.$\left(  \ref{rhoYBH}%
\right)  $. Plugging $\rho_{3}\left(  r\right)  $ into the first EFE, leads to
the following shape function%
\begin{equation}
b(r)=r_{0}+\kappa\left(  b_{3}\left(  r\right)  \exp\left(  -\mu\left(
r-r_{0}\right)  \right)  -b_{3}\left(  r_{0}\right)  \right)  \label{b(r)3}%
\end{equation}
where we have defined%
\begin{equation}
b_{3}(r)=\frac{2C\sqrt{\pi}}{\sqrt{\lambda}}\tan^{-1}\left(  \sqrt{\pi\lambda
}r\right)  -C\frac{\ln\left(  \pi\lambda r^{2}+1\right)  }{\lambda r}.
\end{equation}
Note that when $\mu\rightarrow0$, we recover the energy density in Eq.$\left(
\ref{rho3}\right)  $ and the shape function is simply described by%
\begin{align}
b(r)  &  =r_{0}+\frac{\kappa C}{\lambda}\left(  \frac{\ln\left(  \pi\lambda
r_{0}^{2}+1\right)  }{r_{0}}-\frac{\ln\left(  \pi\lambda r^{2}+1\right)  }%
{r}\right) \nonumber\\
&  +\frac{2\kappa C\sqrt{\pi}}{\sqrt{\lambda}}\left(  \tan^{-1}\left(
\sqrt{\pi\lambda}r\right)  -\tan^{-1}\left(  \sqrt{\pi\lambda}r_{0}\right)
\right)  .
\end{align}
Note also that even for this profile when $\lambda\rightarrow\infty$,
$b(r)=r_{0}$. The shape function $\left(
\ref{b(r)3}\right)$ is such that%
\begin{equation}
b(r)\underset{r\rightarrow\infty}{\rightarrow}r_{0}-b_{3}\left(  r_{0}\right)
=b_{R,\infty}%
\end{equation}
and to avoid that $b_{R,\infty}$ be negative, we need to impose%
\begin{equation}
C\leq\frac{\lambda r_{0}^{2}}{\kappa\left(  2r_{0}\sqrt{\lambda}%
\arctan\!\left(  r_{0}\sqrt{\pi\lambda}\right)  \sqrt{\pi}-\,\ln\!\left(
\pi\lambda r_{0}^{2}+1\right)  \right)  }. \label{CR}%
\end{equation}
To establish the compatibility of $\left(  \ref{CR}\right)  $ with the
flare-out condition, one gets%
\begin{equation}
b^{\prime}(r_{0})=\kappa C\left(  \frac{\ln\!\left(  \pi\lambda r_{0}%
^{2}+1\right)  }{\lambda r_{0}^{2}}\left(  1+\mu r_{0}\right)  -2\mu
\sqrt{\frac{\pi}{\lambda}}\arctan\!\left(  r_{0}\sqrt{\pi\lambda}\right)
\right)  <1 \label{b'(r0)R}%
\end{equation}
and plugging the saturated value of $C$ into $\left(  \ref{b'(r0)R}\right)  $,
we find%
\begin{equation}
b^{\prime}(r_{0})=\frac{\sqrt{\lambda}\ln\!\left(  \pi\lambda r_{0}%
^{2}+1\right)  \left(  1+\mu r_{0}\right)  -2\mu\lambda r_{0}^{2}\sqrt{\pi
}\arctan\!\left(  r_{0}\sqrt{\pi\lambda}\right)  }{2r_{0}\lambda
\arctan\!\left(  r_{0}\sqrt{\pi\lambda}\right)  \sqrt{\pi}-\,\sqrt{\lambda}%
\ln\!\left(  \pi\lambda r_{0}^{2}+1\right)  }<1,
\end{equation}
which is always satisfied. With the help of the relationship $\left(  \ref{or(r)}\right)  $, ZTF can be imposed. For this case we obtain%
\begin{align}
\omega_{r}\left(  r\right)   &  =-\frac{r\left(  \sqrt{\lambda}\,\ln\!\left(
\pi\lambda r_{0}^{2}+1\right)  C\kappa-r_{0}\left(  2\lambda\arctan\!\left(
r_{0}\sqrt{\pi\lambda}\right)  \sqrt{\pi}\,\kappa C-\lambda^{\frac{3}{2}}%
r_{0}\right)  \right)  \exp\left(  \mu\left(  r-r_{0}\right)  \right)  }%
{r_{0}\kappa CD_{R}\left(  r\right)  }\nonumber\\
&  +\frac{-2r_{0}r\lambda C\kappa\arctan\!\left(  \sqrt{\pi\lambda}r\right)
\sqrt{\pi}+r_{0}\sqrt{\lambda}\kappa C\ln\!\left(  \pi\lambda r^{2}+1\right)
}{r_{0}\kappa CD_{R}\left(  r\right)  }, \label{or(r)R}%
\end{align}
where%
\begin{equation}
D_{R}\left(  r\right)  =\sqrt{\lambda}\,\left(  r\mu+1\right)  \ln\!\left(
\pi\lambda r^{2}+1\right)  -2\arctan\!\left(  \sqrt{\pi\lambda}r\right)
\sqrt{\pi}\lambda\mu r^{2}.
\end{equation}
It is immediate to see that for $r\rightarrow\infty$, $\omega_{r}\left(
r\right)  \rightarrow\infty$. One realizes that the saturated value of $C$,
described by the inequality $\left(  \ref{CR}\right)  $ eliminates such
divergence. Indeed, one finds that $\omega_{r}\left(  r\right)  $ becomes%
\begin{align}
\omega_{r}\left(  r\right)   &  =\frac{\lambda}{D_{R}\left(  r\right)
D_{1,R}\left(  r_{0}\right)  }\left(  \left(  2\sqrt{\pi}\arctan\!\left(
\sqrt{\pi\lambda}r\right)  \sqrt{\lambda}r-\ln\!\left(  \pi\lambda
r^{2}+1\right)  \right)  \ln\!\left(  \pi\lambda r_{0}^{2}+1\right)  \right.
\nonumber\\
&  \left.  \left(  2\ln\!\left(  \pi\lambda r^{2}+1\right)  r_{0}\sqrt
{\lambda\pi}-4\pi\arctan\!\left(  \sqrt{\pi\lambda}r\right)  \lambda
rr_{0}\right)  \arctan\!\left(  \sqrt{\pi\lambda}r_{0}\right)  \right)  ,
\end{align}
where we have defined%
\begin{equation}
D_{1,R}\left(  r_{0}\right)  =2\sqrt{\pi}\tan^{-1}\!\left(  \sqrt{\pi\lambda
}r_{0}\right)  \lambda r_{0}-\sqrt{\lambda}\ln\!\left(  \pi\lambda r_{0}%
^{2}+1\right)  .
\end{equation}
On the throat, we obtain%
\begin{align}
\omega_{r}\left(  r_{0}\right)   &  =\frac{\lambda}{D_{R}\left(  r_{0}\right)
D_{1,R}\left(  r_{0}\right)  }\left(  \left(  2\sqrt{\pi}\arctan\!\left(
\sqrt{\pi\lambda}r_{0}\right)  \sqrt{\lambda}r_{0}-\ln\!\left(  \pi\lambda
r_{0}^{2}+1\right)  \right)  \ln\!\left(  \pi\lambda r_{0}^{2}+1\right)
\right. \nonumber\\
&  \left.  \left(  2\ln\!\left(  \pi\lambda r_{0}^{2}+1\right)  r_{0}%
\sqrt{\lambda\pi}-4\pi\arctan\!\left(  \sqrt{\pi\lambda}r_{0}\right)  \lambda
r_{0}^{2}\right)  \arctan\!\left(  \sqrt{\pi\lambda}r_{0}\right)  \right)  .
\end{align}
For $r\rightarrow\infty$, the behaviour of $\omega_{r}\left(  r\right)  $ is%
\begin{equation}
\frac{1}{\mu r}+\!O\left(  \frac{1}{r^{2}}\right)  .
\end{equation}
When we plug the saturated value of $C$, described by Eq.$\left(
\ref{CR}\right)  $ into the shape function $\left(  \ref{b(r)3}\right)  $, one
finds%
\begin{equation}
b(r)=\frac{r_{0}^{2}}{D_{1,R}\left(  r_{0}\right)  r}\left(  2\sqrt{\pi
}\arctan\!\left(  \sqrt{\pi\lambda}r\right)  \lambda r-\ln\left(  \pi\lambda
r^{2}+1\right)  \sqrt{\lambda}\right)  \exp\left(  -\mu\left(  r-r_{0}\right)
\right)  . \label{b(r)RF}%
\end{equation}
We will use the profile $\left(  \ref{b(r)RF}\right)  $ to rearrange $\rho
_{3}\left(  r\right)  $ in the following form%
\begin{equation}
\rho(r)=\frac{r_{0}^{2}}{\kappa r^{4}D_{1,R}\left(  r_{0}\right)  }\left(
\sqrt{\lambda}\,\left(  \mu r+1\right)  \ln\left(  \pi\lambda r^{2}+1\right)
-2\sqrt{\pi}\arctan\!\left(  \sqrt{\pi}\,r\sqrt{\lambda}\right)  \lambda\mu
r^{2}\right)  \exp\left(  -\mu\left(  r-r_{0}\right)  \right)  ,
\end{equation}
where we have used the first EFE. In dimensionless variables, this can also be written in the form
\begin{equation}
\kappa r_0^2\rho(r)=\frac{\left(\left(zx+1\right)  \ln\left(  \pi yx^2+1\right)
-2\sqrt{\pi y}zx^2\arctan\!\left(  \sqrt{\pi y}x\right)\right)}{x^{4}\left(2\sqrt{\pi y}\arctan\left(\sqrt{\pi y}\right)-\ln\left(1+\pi y\right)\right)}  \exp\left(  -z\left(x-1\right)  \right) \ .
\end{equation}
On the other hand the radial pressure
$p_{r}\left(  r\right)  $ becomes%
\begin{equation}
p_{r}\left(  r\right)  =-\frac{r_{0}^{2}}{\kappa r^{4}D_{1,R}\left(
r_{0}\right)  }\left(  2\sqrt{\pi}\arctan\!\left(  \sqrt{\pi\lambda}r\right)
\lambda r-\ln\left(  \pi\lambda r^{2}+1\right)  \sqrt{\lambda}\right)
\exp\left(  -\mu\left(  r-r_{0}\right)  \right)  ,
\end{equation}
where we have used the second EFE. Finally the transverse pressure simply
becomes%
\begin{equation}
p_{t}\left(  r\right)  =\frac{r_{0}^{2}\exp\left(  -\mu\left(  r-r_{0}\right)
\right)  }{2D_{1,R}\left(  r_{0}\right)  r^{7}\kappa^{2}}\left(  \sqrt
{\lambda}\left(  r^{2}\mu^{2}+5\mu r+5\right)  \ln\left(  \pi\lambda
r^{2}+1\right)  -2\lambda\left(  \mu\sqrt{\pi}\left(  \mu r+3\right)
\tan^{-1}\!\left(  \sqrt{\pi\lambda}r\right)  +\sqrt{\lambda}\pi\right)
r^{2}\right)
\end{equation}
One can see that the region of positive energy density is even smaller than in the previous case:
\begin{equation}%
\begin{array}
[c]{cccc}%
x=r/r_{0} & y=\lambda r_{0}^{2} & z=\mu r_{0} & \rho_{3}\left(  r\right)  >0\\
- & 2 & 2 & -\\
- & 1 & 1 & -\\
1<x<1.285 & 1/2 & 1/2 & \surd\\
1<x<2.273 & 1/4 & 1/4 & \surd\\
1<x<8.323 & 1/20 & 1/20 & \surd\\
1<x<2 & 1/2 & 1/4 & \surd\\
1<x<1.437 & 1/4 & 1/2 & \surd
\end{array}
.
\end{equation}
\section{Mixed Energy Density}

\label{ch6}The last energy density profile we are going to modify is the
following%
\begin{equation}
\rho\left(  r\right)  =\frac{3C_{M}^{2}}{8\pi^{2}}\left[  \frac{\pi}{r^{2}%
}-\pi^{2}\lambda\ln\left(  1+\frac{1}{\pi\lambda r^{2}}\right)  \right]
.\label{rho4}%
\end{equation}
Like in section \ref{ch5} and \ref{ch6}, we are going to consider the
following profile%
\begin{equation}
\rho_{4}\left(  r\right)  =\frac{3C_{M}^{2}}{8\pi^{2}r^{2}}\left[  \pi
\exp\left(  -\mu\left(  r-r_{0}\right)  \right)  +\frac{d}{dr}\left(
b_{4}\left(  r\right)  \exp\left(  -\mu\left(  r-r_{0}\right)  \right)
\right)  \right]  ,\label{rho4(r)}%
\end{equation}
where $b_{4}\left(  r\right)  $ is such that%
\begin{equation}
\frac{d}{dr}b_{4}\left(  r\right)  =-\pi^{2}\lambda r^{2}\ln\left(  1+\frac
{1}{\pi\lambda r^{2}}\right)  ,
\end{equation}
with%
\begin{equation}
b_{4}\left(  r\right)  =\frac{2}{3}\sqrt{\frac{\pi}{\lambda}}\arctan\!\left(
\sqrt{\pi\lambda}r\right)  -\frac{r^{3}\lambda\pi^{2}}{3}\left(  \ln\!\left(
\frac{\pi\lambda\,r^{2}+1}{\lambda\pi r^{2}}\right)  \right)  -\frac{2\pi}%
{3}r.
\end{equation}
The shape function becomes%
\begin{equation}
b(r)=r_{0}+\frac{3\kappa C_{M}^{2}}{8\pi^{2}\mu}\left[  \pi\left(
1-\exp\left(  -\mu\left(  r-r_{0}\right)  \right)  \right)  -\mu\left(
b_{4}\left(  r\right)  \exp\left(  -\mu\left(  r-r_{0}\right)  \right)
-b_{4}\left(  r_{0}\right)  \right)  \right]  .\label{b(r)M}%
\end{equation}
Note that when $\mu\rightarrow0$, we recover the energy density in Eq.$\left(
\ref{rho4}\right)  $ and the shape function reduces to%
\begin{equation}
b(r)=r_{0}+\frac{\kappa C_{M}^{2}}{8}\left[  \pi r_{0}^{3}\lambda\ln\!\left(
\frac{1+\pi\lambda r_{0}^{2}}{\lambda\pi r_{0}^{2}}\right)  +\frac{2}%
{\sqrt{\pi\lambda}}\left(  \arctan\!\left(  \sqrt{\pi\lambda}r\right)
-\arctan\!\left(  \sqrt{\pi\lambda}r_{0}\right)  \right)  -r^{3}\lambda
\pi\left(  \ln\!\left(  \frac{\pi\lambda\,r^{2}+1}{\lambda\pi r^{2}}\right)
\right)  +\left(  r-r_{0}\right)  \right]  .
\end{equation}
Note also that $C_{M}$ has dimensions $\left[  L^{-1}\right]  $, $\lambda$ has
dimensions $\left[  L^{-2}\right]  $ while $\mu$ has dimensions $\left[
L^{-1}\right]  $. $C_{M}$, $\lambda$ and $\mu$ are positive. For
$r\rightarrow\infty$ and $\lambda\rightarrow\infty$, $\rho_{4}\left(
r\right)  \rightarrow0$ as well as for $\mu\rightarrow\infty$. However, we
have different asymptotic behaviors depending on which parameter is going to
infinity. For instance, when $\mu\rightarrow\infty$ and $\lambda$ fixed, we
obtain $\rho_{4}\left(  r\right)  \rightarrow0$ corresponding to%
\begin{equation}
b(r)\simeq r_{0}+\frac{3\kappa C_{M}^{2}}{8\pi^{2}}b_{4}\left(  r_{0}\right)
.
\end{equation}
On the other hand, for $\lambda\rightarrow\infty$ and $\mu$ fixed, one finds
that the shape function $\left(  \ref{b(r)M}\right)  $ becomes%
\begin{equation}
b(r)=r_{0}+\frac{3\kappa C_{M}^{2}}{8\pi\mu}\left[  1+\mu r_{0}-\left(  1+\mu
r\right)  \exp\left(  -\mu\left(  r-r_{0}\right)  \right)  \right]  ,
\end{equation}
corresponding to%
\begin{equation}
\rho_{4}\left(  r\right)  \simeq\frac{3\kappa C_{M}^{2}\mu\exp\left( -\mu\left(
r-r_{0}\right)  \right)  }{8\pi r}.
\end{equation}
It is interesting to observe that for $\lambda$, $\mu\rightarrow\infty$ we
obtain%
\begin{equation}
b(r)\simeq r_{0}+\frac{3\kappa C_{M}^{2}}{8\pi^{2}}.
\end{equation}
For $\lambda\rightarrow0$, $\rho_{4}\left(  r\right)  $ reduces to
\begin{equation}
\bar{\rho}_{4}\left(  r\right)  =\frac{3C_{M}^{2}\exp\left(  -\mu\left(
r-r_{0}\right)  \right)  }{8\pi r^{2}},
\end{equation}
namely the YBH HDE energy density profile up to a numerical factor, whose
solution is%
\begin{equation}
b(r)=r_{0}+\frac{3\kappa C_{M}^{2}}{8\pi\mu}\left(  1-\exp\left(  -\mu\left(
r-r_{0}\right)  \right)  \right)  .
\end{equation}
Even for the mixed case, we would like to see if it is possible to impose ZTF.
To this purpose, we compute Eq.$\left(  \ref{or(r)}\right)  $ and we obtain%
\begin{equation}
\omega_{r}\left(  r\right)  =\frac{N_{4a}\left(  r\right)  \exp\left(
\mu\left(  r-r_{0}\right)  \right)  +N_{4b}\left(  r\right)  }{D_{4}\left(
r\right)  },
\end{equation}
where%
\begin{align}
N_{4a}\left(  r\right)   &  =\kappa C^{2}\left(  2\mu\arctan\!\left(
r_{0}\sqrt{\pi\lambda}\right)  -\sqrt{\pi\lambda}\left(  \pi\mu\lambda
r_{0}^{3}\ln\!\left(  \frac{\pi\lambda r_{0}^{2}+1}{\lambda\pi r_{0}^{2}%
}\right)  +\left(  2\mu r_{0}+3\right)  \right)  \right)  -4r_{0}\mu\,\pi
\sqrt{\pi\lambda},\\
N_{4b}\left(  r\right)   &  =\kappa C^{2}\left(  \sqrt{\pi\lambda}\left(
\pi\mu\lambda r^{3}\ln\!\left(  \frac{\pi\lambda r^{2}+1}{\lambda\pi r^{2}%
}\right)  +\left(  2\mu r+3\right)  \right)  -2\arctan\!\left(  \sqrt
{\pi\lambda}r\right)  \mu\right)  ,\\
D_{4}\left(  r\right)   &  =\mu r\kappa C^{2}\left(  \pi^{\frac{3}{2}}%
\lambda^{\frac{3}{2}}r^{2}\left(  \mu r-3\right)  \ln\!\left(  \frac
{\pi\lambda r^{2}+1}{\pi\lambda r^{2}}\right)  -2\arctan\!\left(  \sqrt
{\pi\lambda}r\right)  \mu+\sqrt{\lambda\pi}\left(  2\mu r+3\right)  \right)  .
\end{align}
As we can see, also for this case there appears a divergence for
$r\rightarrow\infty$ which can be eliminated if $N_{4a}\left(  r\right)  $
vanishes. This is possible if we assume that%
\begin{equation}
C=\sqrt{\frac{4r_{0}\mu\,\pi\sqrt{\pi\lambda}}{\kappa\left(  2\mu
\arctan\!\left(  r_{0}\sqrt{\pi\lambda}\right)  -\sqrt{\pi\lambda}\left(
\pi\mu\lambda r_{0}^{3}\ln\!\left(  \left(  \pi\lambda r_{0}^{2}+1\right)
/\left(  \lambda\pi r_{0}^{2}\right)  \right)  +\left(  2\mu r_{0}+3\right)
\right)  \right)  }}.\label{CM}%
\end{equation}
This means that the ZTF can be imposed and also the radial and transverse
pressures can be easily computed. To further proceed, we substitute the value
of $C$ into the shape function $\left(  \ref{b(r)M}\right)  $ to obtain%
\begin{equation}
b(r)=\frac{r_{0}\exp\left(  -\mu\left(  r-r_{0}\right)  \right)  N\left(
r\right)  }{N\left(  r_{0}\right)  },
\end{equation}
where we have defined%
\begin{equation}
N\left(  r\right)  =2\mu\arctan\!\left(  \sqrt{\pi\lambda}r\right)  -\mu
r^{3}\,\sqrt{\left(  \pi\lambda\right)  ^{3}}\ln\!\left(  \left(  \pi\lambda
r^{2}+1\right)  /\left(  \pi\lambda r^{2}\right)  \right)  -\sqrt{\pi\lambda
}\left(  2\mu r+3\right)  .
\end{equation}
In terms of $N\left(r \right )$, the energy density can be cast into the form
\begin{equation}
\rho_4 \ = \ \frac{r_0\left(N'\left(r \right )-\mu N\left(r \right )\right)}{\kappa r^2N(r_0)}\exp\left(  -\mu\left(  r-r_{0}\right)  \right) \ .
\end{equation}
To complete the calculation of the SET, we compute the radial pressure%
\begin{equation}
p_{r}\left(  r\right)  =-\frac{b(r)}{\kappa r^{3}}=-\frac{r_{0}\exp\left(
-\mu\left(  r-r_{0}\right)  \right)  N\left(  r\right)  }{\kappa r^{3}N\left(
r_{0}\right)  }%
\end{equation}
and the transverse pressure%
\begin{equation}
p_{t}\left(  r\right)  =\frac{b\left(  r\right)  -b^{\prime}\left(  r\right)
r}{2\kappa r^{3}}={
 \frac{r_0 e^{-\mu(r-r_0)}}{2\kappa r^3 N(r_0)}
\left[(1+\mu r)N(r)-rN'(r)\right]
},
\end{equation}
where%
\begin{equation}
N^{\prime}\left(  r\right)  =3\ln\!\left(  \frac{\pi\lambda r^{2}}{\pi\lambda
r^{2}+1}\right)  \sqrt{\left(  \lambda\pi\right)  ^{3}}r^{2}\mu.
\end{equation}

Surprisingly we can find that $\rho_{4}\left(  r\right)$ is generally negative for all values considered in the table below:
\begin{equation}%
\begin{array}
[c]{cccc}%
x=r/r_{0} & y=\lambda r_{0}^{2} & z=\mu r_{0} & \rho_{3}\left(  r\right)  >0\\
- & 2 & 2 & -\\
- & 1 & 1 & -\\
- & 1/2 & 1/2 & -\\
- & 1/4 & 1/4 & -\\
- & 1/20 & 1/20 & -\\
- & 1/2 & 1/4 & -\\
- & 1/4 & 1/2 & -
\end{array}
.
\end{equation}
\section{Conclusion}
\label{ch7}In this paper, we have extended the analysis performed in Ref.%
\cite{RGPC} involving different energy density profiles inspired by
holographic dark energies as possible sources needed to have traversable
wormhole solutions. The extension we have made consists in an introduction
of a Yukawa term in such a way to restrict the effect of the source in a
limited range. The profiles we have explored are: the Bekenstein-Hawking
HDE, the Moradpour energy density, the Standard Renyi HDE and the Mixed
Energy Density. All these profiles have a positive energy density, but the
NEC must be violated to obtain a TW. Thus we are forced to introduce an EoS
of the form $\left( \ref{or(r)}\right) $ allowing the following inequality%
\begin{equation}
\rho \left( r\right) +p_{r}\left( r\right) =\left( 1+\omega _{r}\left(
r\right) \right) \rho (r)\leq 0
\end{equation}%
with $\omega _{r}\left( r\right) <-1$. This means that our energy density
profiles are of the \textquotedblleft \textit{phantom}\textquotedblright\
type. With such an EoS, we have tried to impose ZTF, that it means that $%
\Phi (r)$ can assume a constant value or it can be vanishing. The purpose of
this paper is to verify if all the above mentioned profiles drastically
change behavior if a Yukawa term is included. The original HDE profiles
examined in Ref.\cite{RGPC} had as a side effect a divergent $\omega
_{r}\left( r\right) $ when $r\rightarrow \infty $. This unpleasant behavior
has been solved with the modification of the $\omega _{r}\left( r\right) $
function in such a way that the behavior at infinity is convergent without
consequences on the behavior on the throat. In the Yukawa distorted cases
only the YBH-HDE profile has requested such a modification. This procedure
has been also used in Ref.\cite{RGFLKZ} where a positive Dark Matter profile
has been considered. Nevertheless the price to pay is that the positivity of
the source is restricted in a well defined small space-time region with the
exception of the Yukawa Mixed Energy Density which is always negative. In
other words, the Yukawa distortion necessarily introduces a certain amount
of negative energy density. Actually, the negativity of energy density is
not directly caused by the Yukawa modification itself. Instead, it is the
result of our desire to have an integral in closed form, and thus the
introduction of extra negative terms that start to dominate at larger
distances. Moreover such extra negative terms are the consequence of the
fact that the Yukawa distortion appears under a derivative operation. This
is the reason of why only the YBH-HDE profile is the only profile that
remains positive. This means that the NEC is violated entirely by the radial
pressure. It is interesting to observe that the $a=0$ case is the one that
looks like a pure Yukawa profile, even if the denominator has a different
exponent. It is also remarkable that $a=0$ has a total gravitational energy $%
E_{G}\left( r\right) $ finite $\forall r\in \left[ r_{0},+\infty \right) $.
On the other hand, at this stage of the investigation we do not know the
physical reason of why only the Yukawa Mixed Energy Density behaves in this
way.

\section*{Acknowledgments}

P. Channuie is partially supported by the Thailand National Science, Research and Innovation Fund (TSRF) via PMU-B with grant No.B39G690007.


\end{document}